\documentclass{article}
\usepackage{amsmath}
\usepackage{cite}
\usepackage{graphicx}
\usepackage{dcolumn}

\begin{document}

\date{}
\title{On the exact solutions of a two-dimensional hydrogen atom in a constant
magnetic field}
\author{Francisco M. Fern\'{a}ndez\thanks{%
fernande@quimica.unlp.edu.ar} \\
INIFTA, DQT, Sucursal 4, C. C. 16, \\
1900 La Plata, Argentina}
\maketitle

\begin{abstract}
We discuss the exact polynomial solutions for the two-dimensional
hydrogen atom in a constant magnetic field already studied earlier
by other authors. In order to provide a suitable meaning for such
solutions we compare them with numerical results provided by the
Rayleigh-Ritz method.
\end{abstract}

\section{Introduction}

\label{sec:intro}

There are many exactly-solvable quantum-mechanical models. The most popular
an useful ones may be the harmonic oscillator and the hydrogen atom that are
discussed in any textbook on quantum mechanics\cite{CDL77} and quantum
chemistry\cite{P68}. Several years ago, Flessas and collaborators\cite
{F79,FD80,F81,F81b,FW81} discovered that some quantum-mechanical models that
are not exactly solvable admit some exact solutions for particular values of
the model parameters. This kind of problems are now known as quasi solvable
(QS) or conditionally solvable and have been widely studied\cite{T16}.
Several researchers thought that the QS solutions are the exact solutions of
the quantum-mechanical models and derived incorrect conclusions from them as
discussed elsewhere\cite{F21,AF21}. In fact, such QS solutions are of scarce
utility if one is not able to connect them with the actual solutions of the
quantum-mechanical models\cite{F21,AF21} (see, for example, the enlightening
papers by Child et al\cite{CDW00} and Le et al\cite{LHL17}).

Recently, Bildstein and Grabowski\cite{BG25} proposed a
generalization of the Frobenius method and applied it to
quantum-mechanical models with polynomial potentials in one, two
and three dimensions. They obtained analytical solutions for most
of them. In particular, they considered in detail the
two-dimensional hydrogen atom in a constant magnetic field already
studied by Le et al\cite{LHL17} and somewhat earlier by Taut\cite
{T95}. In particular, Turbiner and Escobar Ruiz discussed the
existence of a hidden algebra. The purpose of this paper is the
interpretation of the exact QS solutions for this model.

In section~\ref{sec:model} we outline some relevant features of the Schr\"{o}%
dinger equation for the two-dimensional hydrogen atom in a constant magnetic
field. In section~\ref{sec:exact_poly} we derive some exact analytical
solutions by means of the Frobenius method and a simple truncation rule
already used in earlier papers\cite{F21,AF21}. In order to provide a sound
interpretation of the analytical QS results, in section~\ref{sec:RRM} we
compare them with numerical results coming from the well known Rayleigh-Ritz
method (RRM)\cite{P68} that yields increasingly accurate upper bounds to the
actual eigenvalues of the Schr\"{o}dinger equation\cite{M33,F25a}. Finally,
in section~\ref{sec:conclusions} summarize the main results and draw
conclusions.

\section{The model}

\label{sec:model}

The Schr\"{o}dinger equation for the two-dimensional hydrogen atom in a
constant magnetic field is separable in polar coordinates and the resulting
radial eigenvalue equation in atomic units is\cite{LHL17}
\begin{equation}
\left[ -\frac{1}{2}\left( \partial _{r}^{2}+\frac{1}{r}\partial _{r}\right) +%
\frac{m^{2}}{2r^{2}}+\frac{\gamma ^{2}}{8}r^{2}-\frac{Z}{r}\right]
R(r)=WR(r),\;W=E-\frac{\gamma m}{2},  \label{eq:eigen_eq}
\end{equation}
where $E$ is the energy and $m=0,\pm 1,\pm 2,\ldots $ is the magnetic
quantum number. This equation is a particular case of a more general one
studied in our earlier paper\cite{F21}. Bound states are solutions of this
eigenvalue equation that are square integrable
\begin{equation}
\int_{0}^{\infty }R(r)^{2}r\,dr<\infty .  \label{eq:bound_cond}
\end{equation}
Such solutions are possible only for discrete values of $W$ that we may
denote $W_{\nu s}\left( Z,\gamma \right) $, where $s=|m|$ and $\nu
=0,1,\ldots $ is the radial quantum number (the number of nodes of $R(r)$ in
the interval $0<r<\infty $). The corresponding square-integrable solutions
are $R_{\nu s}(r)$. It is well known that $W_{\nu s}>W_{\nu ^{\prime }s}$ if
$\nu >\nu ^{\prime }$ and $W_{\nu s}>W_{\nu s^{\prime }}$ if $s>s^{\prime }$.

It is clear that if $\gamma ^{2}>0$ then there are bound states for all $%
-\infty <Z<\infty $. From now on we only consider $Z>0$ that is consistent
with the physical model. According to the Hellmann-Feynman theorem (HFT)\cite
{G32,F39}
\begin{equation}
\frac{\partial W_{\nu s}}{\partial Z}=-\left\langle \frac{1}{r}\right\rangle
<0,\;\frac{\partial W_{\nu s}}{\partial \gamma }=\frac{\gamma }{4}%
\left\langle r^{2}\right\rangle >0,\;\gamma >0,  \label{eq:HFT}
\end{equation}
we conclude that the eigenvalues decrease with $Z$ and increase with $\gamma
$. For $\gamma =0$ all the eigenvalues are negative
\begin{equation}
W_{\nu s}\left( Z,0\right) =-\frac{2Z^{2}}{\left( 2\nu +2s+1\right) ^{2}},
\label{eq:E_HA}
\end{equation}
while for $Z=0$ they are positive
\begin{equation}
W_{\nu s}\left( 0,\gamma \right) =\frac{1}{2}\left( 2\nu +s+1\right) \gamma .
\label{eq:W_HO}
\end{equation}
Therefore, as $\gamma $ increases the eigenvalues $W_{\nu s}\left( Z,\gamma
\right) $ become positive and there exist $\gamma =\gamma _{\nu s}^{c}$ such
that $W_{\nu s}=0$. These critical values of $\gamma $ satisfy $\gamma _{\nu
s}^{c}<\gamma _{\nu ^{\prime }s}^{c}$ if $\nu >\nu ^{\prime }$ and $\gamma
_{\nu s}^{c}<\gamma _{\nu s^{\prime }}^{c}$ if $s>s^{\prime }$.

By means of a suitable change of variables\cite{F20} we may set either $%
|Z|=1 $ or $\gamma =1$ (preferably $\gamma =2$ in order to derive
an equation similar to that studied in our earlier
paper\cite{F21}). However, here we will keep both model parameters
in order to compare present results with those of Bildstein and
Grabowski\cite{BG25}.

The analytical results outlined above may appear to be rather trivial and
well known. We decided to show them here because most of them have been
overlooked by the researchers who drawn incorrect conclusions from the exact
polynomial solutions of QS models as discussed in our earlier paper\cite{F21}
(se allso\cite{AF21} )

\section{Exact polynomial solutions}

\label{sec:exact_poly}

In this section we will derive exact solutions to the eigenvalue equation (%
\ref{eq:eigen_eq}) by means of the standard Frobenius method used in our
earlier paper\cite{F21} (in fact, Taut\cite{T95} had used it several years
before). On taking into account the asymptotic behaviour of $R(r)$ at origin
$R(r)\sim r^{s}$ and at infinity $R(r)\sim e^{-\gamma r^{2}/4}$, we propose
an ansatz of the form
\begin{equation}
R(r)=r^{s}e^{-\gamma r^{2}/4}\sum_{j=0}^{\infty }c_{j}r^{j}.  \label{eq:R(r)}
\end{equation}
The expansion coefficients $c_{j}$ satisfy the three-term recurrence
relation (TTRR)
\begin{eqnarray}
c_{j+2} &=&A_{j}c_{j+1}+B_{j}c_{j},\;j=0,1,\ldots ,  \nonumber \\
A_{j} &=&-\frac{2Z}{\left( j+2\right) \left( j+2\left( s+1\right) \right) }%
,\;B_{j}=\frac{\gamma \left( j+s+1\right) -2W}{\left( j+2\right) \left(
j+2\left( s+1\right) \right) }.  \label{eq:TTRR}
\end{eqnarray}

In order to obtain an exact polynomial solution of degree $n$ we require
that $c_{n}\neq 0$ and $c_{n+1}=c_{n+2}=0$, $n=0,1,\ldots $, so that $%
c_{j}=0 $ for all $j>n.$ These conditions lead to $B_{n}=0$ from which we
obtain
\begin{equation}
W=W_{s}^{(n)}=\frac{\gamma \left( n+s+1\right) }{2},  \label{eq:W^(n)_s}
\end{equation}
and
\begin{equation}
B_{j}=\frac{\gamma \left( j-n\right) }{\left( j+2\right) \left( j+2\left(
s+1\right) \right) }.  \label{eq:B_jn}
\end{equation}
It only remains to solve the condition $c_{n+1}=0$ for either $Z$ or $\gamma
$. Here, we follow Bildstein and Graboski\cite{BG25} (who had followed Le at
al\cite{LHL17}) and solve for $\gamma $ though solving for $Z$ is perhaps
more convenient\cite{F21,AF21}. In any case, the exact polynomial solutions
may be written as
\begin{equation}
R_{s}^{(n)}(r)=r^{s}e^{-\gamma r^{2}/4}\sum_{j=0}^{n}c_{js}^{(n)}r^{j},
\label{eq:R^(n)}
\end{equation}
but for simplicity we will write $c_{j}$ for the expansion coefficients in
what follows.

Before proceeding with the calculations note that $W_{s}^{(n)}$ is positive
for all values of $n$, $s$ and $\gamma $ so that the exact polynomial
solutions cannot provide information on the whole spectrum of the problem.
Besides, $n$ is not the radial quantum number as shown below. Many
researchers wrongly interpreted $W_{s}^{(n)}$ as the spectrum of the problem
and $n$ as the radial quantum number as argued in earlier papers\cite
{F21,AF21}.

When $n=0$ we have
\begin{equation}
c_{1}=-\frac{2Z}{2s+1}.  \label{eq:n=0,c1}
\end{equation}
The only solution for $c_{1}=0$ is $Z=0$ and we obtain a set of exact
solutions for the harmonic oscillator
\begin{equation}
W_{s}^{(0)}=W_{0s}=\frac{\gamma (s+1)}{2}%
,\;R_{s}^{(0)}(r)=R_{0s}(r)=c_{0}r^{s}e^{-\gamma r^{2}/4}.
\label{eq:n=0,W,R}
\end{equation}
Note that $\gamma $ is arbitrary because the harmonic oscillator is exactly
solvable for all values of this parameter.

When $n=1$ the second and third coefficients are
\begin{equation}
c_{1}=-\frac{2Z}{2s+1},\;c_{2}=\frac{Z^{2}}{(s+1)(2s+1)}-\frac{\gamma }{%
4(s+1)},  \label{eq:n=0,c1,c2}
\end{equation}
from which we obtain
\begin{equation}
\gamma =\frac{4Z^{2}}{2s+1},  \label{eq:n=0,gamma}
\end{equation}
and
\begin{equation}
W_{s}^{(1)}=\frac{2(s+2)Z^{2}}{2s+1},\;R_{s}^{(1)}(r)=c_{0}r^{s}e^{-\gamma
r^{2}/4}\left( 1-\frac{2Zr}{2s+1}\right) .  \label{eq:n=1,W,R}
\end{equation}
We appreciate that the truncation method fails to provide the solutions
without nodes for $Z>0$, a fact that was pointed out by Taut\cite{T95} many
years ago.

When $n=2$ the termination condition is
\begin{equation}
c_{3}=\frac{z\left[ \gamma \left( 4s+3\right) -2z^{2}\right] }{3\left(
s+1\right) \left( 2s+1\right) \left( 2s+3\right) }=0.  \label{eq:n=2,c_3}
\end{equation}
When $Z=0$ we obtain the solutions of the harmonic oscillator with one node
\begin{equation}
W_{s}^{(2)}=W_{1s}=\frac{\gamma (s+3)}{2}%
,\;R_{s}^{(2)}(r)=R_{1s}(r)=c_{0}r^{s}e^{-\gamma r^{2}/4}\left[ 1-\frac{%
\gamma r^{2}}{2\left( s+1\right) }\right] .  \label{eq:n=2,W,R,Z=0}
\end{equation}
If we solve equation (\ref{eq:n=2,c_3}) for $\gamma $ we have
\begin{equation}
\gamma =\frac{2Z^{2}}{4s+3},  \label{eq:n=2,gamma}
\end{equation}
and
\begin{equation}
W_{s}^{(2)}=\frac{Z^{2}(s+3)}{4s+3},\;R_{s}^{(2)}(r)=c_{0}r^{s}e^{-\gamma
r^{2}/4}\left[ 1-\frac{2Zr}{2s+1}+\frac{2Z^{2}r^{2}}{\left( 2s+1\right)
\left( 4s+3\right) }\right] .  \label{eq:n=2,W,R}
\end{equation}
The truncation method yields a solution $R_{s}^{(2)}(r)$ with two nodes and
fails to provide those with zero and one node. Le et al\cite{LHL17} proved a
most remarkable theorem on the nodes of the exact polynomial solutions.

These results agree with those derived by Bildstein and Grabowski who also
showed expressions for $n=3$. The most important point here is that
Bildstein and Grabowski did not mention that $W_{s}^{(n)}(Z)$ are not the
eigenvalues of the radial equation (\ref{eq:eigen_eq}) which is the reason
why they do not exhibit the correct behaviour with respect to $Z$ (see
equation (\ref{eq:HFT})). The behaviour of $W_{\nu s}(Z)$ with respect to $Z$
has already been discussed in two earlier papers\cite{F21,AF21} and the
limitations of the exact polynomial solutions have been discussed by Taut%
\cite{T95} and Le et al\cite{LHL17}. A suitable interpretation of the QS
results is most important as argued in earlier papers\cite{F21,AF21}.

Here, we derive two general results. First, since $B_{j}$ is a linear
function of $\gamma $ and $A_{j}$ does not depend on this parameter, then if
$c_{j}$ is a polynomial function of $\gamma $ of degree $k$ then $c_{j+2}$
is a polynomial function of this parameter of degree $k+1$. Consequently,
for $n>0$ we obtain a set of roots $\gamma _{s}^{(n,i)}$, $i=1,2,\ldots
,k_{n}$, where $k_{2n}=k_{2n-1}=n$, $n=1,2,\ldots $. This result was already
derived by Le et al\cite{LHL17} in a somewhat different way. The
corresponding QS eigenvalues $W_{s}^{(n,i)}=W_{s}^{(n)}\left( \gamma
_{s}^{(n,i)}\right) $ are associated to the eigenfunctions $R_{s}^{(n,i)}(r)$
with expansion coefficients $c_{js}^{(n,i)}$.

Second, if we substitute $\gamma =\xi Z^{2}$ and $c_{j}=d_{j}Z^{j}$ in the
recurrence relation (\ref{eq:TTRR}) with $B_{j}$ given by (\ref{eq:B_jn}) we
obtain the TTRR
\begin{equation}
d_{j+2}=-\frac{2}{\left( j+2\right) \left( j+2\left( s+1\right) \right) }%
d_{j+1}+\frac{\xi \left( j-n\right) }{\left( j+2\right) \left( j+2\left(
s+1\right) \right) }d_{{j}},  \label{eq:TTRR_d_j}
\end{equation}
that is exactly the TTRR relation (\ref{eq:TTRR}) with $Z=1$. We thus
conclude that $\gamma _{s}^{(n)}(Z)=\gamma _{s}^{(n)}(1)Z^{2}$ and $%
c_{j}(Z)=c_{j}(1)Z^{j}$. It is obvious that we can set $Z=1$ without loss of
generality.

In the next section we provide an interpretation of the QS results as was
done by Le et al\cite{LHL17}. In closing this section, we just mention that $%
A_{j}=0$ for all $j$ when $Z=0$ in which case we obtain the two-term
recurrence relation for the harmonic oscillator $c_{2j+2}=B_{2j}c_{2j}$
because $c_{2j+1}=0$ for all $j$.

\section{Variational calculation}

\label{sec:RRM}

One way of providing a suitable interpretation of the QS results is to
compare them with accurate numerical eigenvalues for an interval of the
chosen model parameter. Le et al\cite{LHL17} resorted to an approach based
on the iterative solution of the secular equation\cite{HPL13} already
proposed earlier by other authors\cite{FMC85}. In this section, we carry out
some numerical calculations based on the RRM\cite{P68} that is known to
provide increasingly tighter upper bounds to the exact eigenvalues\cite
{M33,F25a}. For simplicity, we choose the non-orthogonal basis set
\begin{equation}
u_{is}(r)=e^{i+s}e^{-\gamma r^{2}/4},\;i=0,1,\ldots ,  \label{eq:RRM_basis_1}
\end{equation}
that is suitable for sufficiently large values of $\gamma $. The equations
of the RRM with a non-orthogonal basis set are well known\cite{P68,F24} and
will not be shown here. For concreteness, we restrict our calculations to $%
Z=1$ and $s=0$.

Tables \ref{Tab:RRM1} and \ref{Tab:RRM2} show the convergence of the RRM
eigenvalues for $\gamma =4$ and $\gamma =2/3$, respectively, as the
dimension $N$ of the basis set increases. The RRM yields the exact QS
results $W_{10}=W_{0}^{(1)}=4$ in the former case and $W_{20}=W_{0}^{(2)}=1$
in the latter. This fact was already pointed out by Le et al\cite{LHL17} but
we can show it more clearly in terms of present approach. In fact, the
secular determinants $D_{N}$ for $N=2$ and $N=3$ exhibit the exact results
\begin{eqnarray}
D_{2} &=&\frac{4-\pi }{128}\left( W-4\right) \left[ W-\frac{2\sqrt{2}\left(
\sqrt{\pi }-\sqrt{2}\right) }{\pi -4}\right] ,  \nonumber \\
D_{3} &=&\frac{81}{128}\left( W-1\right) \left[ 9\left( 5\pi -16\right)
W^{2}+6\left( 3\sqrt{3}\pi ^{\frac{3}{2}}-6\pi -10\sqrt{3}\sqrt{\pi }%
+20\right) W\right.  \nonumber \\
&&\left. -2\sqrt{3}\pi ^{\frac{3}{2}}-11\pi +10\sqrt{3}\sqrt{\pi }+24\right]
.  \label{eq:D_2,D_3}
\end{eqnarray}
It also follows from Tables \ref{Tab:RRM1} and \ref{Tab:RRM2} that the
truncation approach is unable to provide the ground state $W_{00}(\gamma )$
as argued in the preceding section and as was pointed out several years ago
by Taut\cite{T95}.

\begin{figure}[tbp]
\begin{center}
\includegraphics[width=9cm]{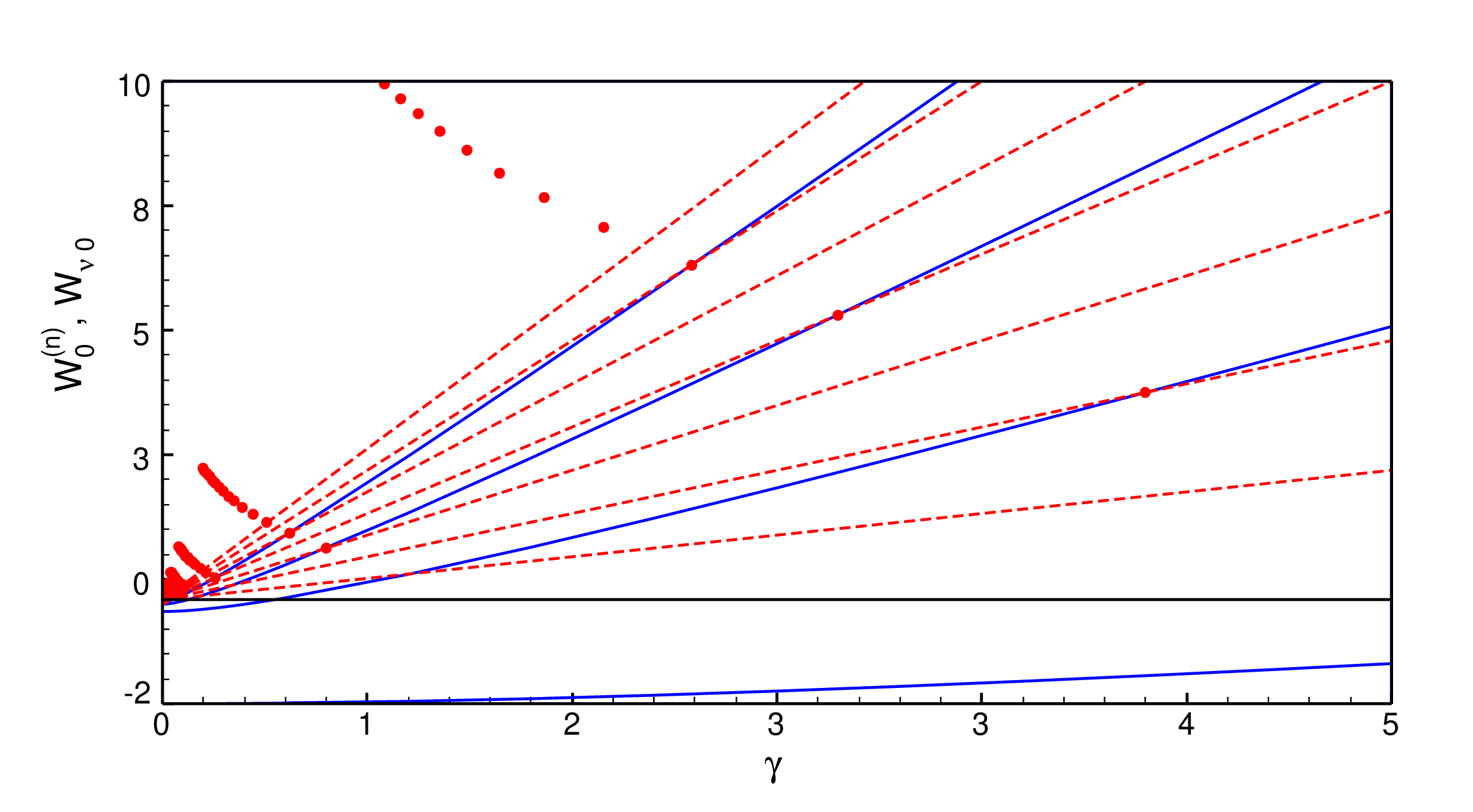}
\end{center}
\caption{RRM eigenvalues $W_{\nu 0}$, $\nu=0,1,2,3$ (blue continuous lines),
TTRR eigenvalues $W^{(n)}_0(\gamma)$, $n=0,1,\ldots,6$ (red dashed lines)
and $W^{(n)}_0\left(\gamma^{(n)}_0\right)$ (red points)}
\label{Fig:Figure1}
\end{figure}

Figure\ref{Fig:Figure1} shows RRM eigenvalues $W_{\nu 0}(\gamma )$, $\nu
=0,1,2,3$, and QS eigenvalues $W_{0}^{(n)}(\gamma )$ for $n=0,1,\ldots ,6$.
We appreciate that every point $\left[ \gamma
_{s}^{(n,i)},W_{0}^{(n,i)}\right] $ marks an intersection between a pair of
curves $W_{\nu 0}(\gamma )$ and $W_{0}^{(n)}(\gamma )$. However, there are
intersections between curves $W_{\nu 0}(\gamma )$ and $W_{0}^{(n)}(\gamma )$
that do not correspond to points $\left[ \gamma
_{s}^{(n,i)},W_{0}^{(n,i)}\right] $. This figure provides a clear
interpretation of the QS results. Le et al\cite{LHL17} resorted to somewhat
different plots for the same purpose. On the other hand, Bildstein and
Grabowski\cite{BG25} did not attempt to provide such interpretation which is
necessary to avoid the mistakes mentioned above.

The rate of convergence of the RRM with the basis set (\ref{eq:RRM_basis_1})
decreases as $\gamma $ decreases. For small values of $\gamma $, say $\gamma
<1$, it is convenient to resort to the alternative basis set
\begin{equation}
v_{is}(r)=e^{i+s}e^{-\alpha r},\;i=0,1,\ldots ,  \label{eq:RRM_basis_2}
\end{equation}
where $\alpha $ is an adjustable parameter. One obtains the greatest rate of
convergence when $\alpha $ is determined variationally but here we just
choose $\alpha =1$ in all our calculations because we do not need high
accuracy.

\begin{figure}[tbp]
\begin{center}
\includegraphics[width=9cm]{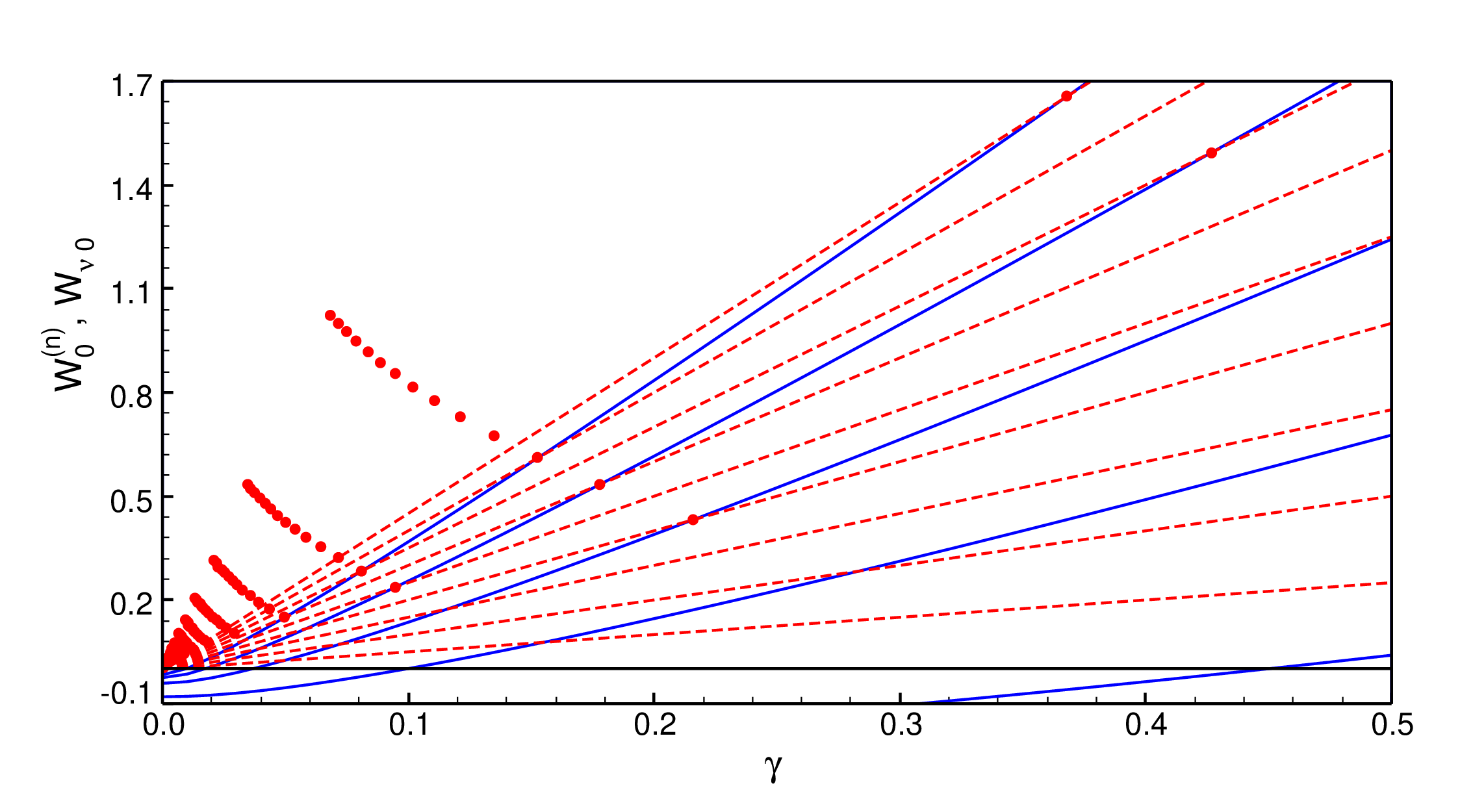}
\end{center}
\caption{RRM eigenvalues $W_{\nu 0}$, $\nu=1,2,\ldots,5$ (blue continuous
lines), TTRR eigenvalues $W^{(n)}_0(\gamma)$, $n=0,1,\ldots,8$ (red dashed
lines) and $W^{(n)}_0\left(\gamma^{(n)}_0\right)$ (red points) }
\label{Fig:Figure2}
\end{figure}

Figure~\ref{Fig:Figure2} shows the QS eigenvalues $W_{0}^{(n)}(\gamma )$, $%
n=0,1,\ldots ,8$, and the RRM ones $W_{\nu 0}$, $\nu =1,2,\ldots ,5$, in an
interval of values of $\gamma $ closer to the origin. The conclusions are
similar to those drawn from Figure~\ref{Fig:Figure1}.

Although at first sight the critical values of $\gamma $ do not appear to be
related with the analysis of the QS results, it is clear that the truncation
approach cannot provide any information about the actual eigenvalue $W_{\nu
s}$ when $\gamma <\gamma _{\nu s}^{c}$. For this reason, we have decided to
calculate some critical values of $\gamma $. Table~\ref{tab:crit} shows
values of $\gamma _{\nu s}^{c}$ for $s=0$ and $\nu =0,1,\ldots ,9$. In order
to obtain them we simply set $W=0$ in the RRM secular equation and solved
for $\gamma $. We also resorted to the Riccati-Pad\'{e} method\cite{FMT89}
in order to obtain more accurate results in some cases. Taking into account
figures \ref{Fig:Figure1} and \ref{Fig:Figure2} and the extremely large
value of $\gamma _{00}^{c}$ we understand why the QS results fail to include
the ground state of the model.

\section{Further comments and conclusions}

\label{sec:conclusions}

Bildstein and Grabowski\cite{BG25} derived some isolated exact solutions of
the eigenvalue equation (\ref{eq:eigen_eq}) by means of a truncation
procedure used earlier by other authors\cite{T95,LHL17,F21,AF21} but did not
try to connect them with the actual eigenvalues and eigenfunctions. The
correct interpretation of the QS solutions is most important in order to
avoid the mistakes discussed elsewhere\cite{F21,AF21}. For this reason we
decided to carry out present analysis which was presented in a way that
slightly differs from those of Taut\cite{T95} and Le et al\cite{LHL17}.

\begin{table}[tbp]
\caption{Lowest RRM eigenvalues for $Z=1$, $s=0$ and $\gamma=4$}
\label{Tab:RRM1}
\begin{center}
\begin{tabular}{llcll}
$N$ & \multicolumn{1}{c}{$\nu=0$} & \multicolumn{1}{c}{$\nu=1$} &
\multicolumn{1}{c}{$\nu=2$} & \multicolumn{1}{c}{$\nu=3$} \\
4 & -1.449885589 & 4 & 8.34525977 & 17.66452696 \\
5 & -1.458156835 & 4 & 8.344361267 & 12.69095166 \\
6 & -1.459389343 & 4 & 8.344349784 & 12.53313314 \\
7 & -1.459560848 & 4 & 8.344349441 & 12.53290257
\end{tabular}
\end{center}
\end{table}

\begin{table}[tbp]
\caption{Lowest RRM eigenvalues for $Z=1$, $s=0$ and $\gamma=\frac{2}{3}$}
\label{Tab:RRM2}
\begin{center}
\begin{tabular}{lllcl}
$N$ & \multicolumn{1}{c}{$\nu=0$} & \multicolumn{1}{c}{$\nu=1$} &
\multicolumn{1}{c}{$\nu=2$} & \multicolumn{1}{c}{$\nu=3$} \\
4 & -1.835656791 & 0.184392123 & 1 & 1.904674543 \\
5 & -1.935195212 & 0.181236739 & 1 & 1.743664442 \\
6 & -1.968131654 & 0.1807724337 & 1 & 1.743410741 \\
7 & -1.976985355 & 0.1807106702 & 1 & 1.743408135
\end{tabular}
\end{center}
\end{table}

\begin{table}[tbp]
\caption{Some critical values of the parameter $\gamma$}
\label{tab:crit}
\begin{center}
\begin{tabular}{ll}
$\nu$ & \multicolumn{1}{c}{$\gamma^c_{\nu 0}$} \\
0 & 9.399451214 \\
1 & 0.4484067794 \\
2 & 0.09870506669 \\
3 & 0.03616422276 \\
4 & 0.01705321756 \\
5 & 0.005668542649 \\
6 & 0.003691507495 \\
7 & 0.002536579450 \\
8 & 0.001817259637 \\
9 & 0.001024706586
\end{tabular}
\end{center}
\end{table}

\end{document}